\documentclass{PoS}

\usepackage{amstext}  
\usepackage{amsfonts}
\usepackage{amssymb} 
\usepackage{amsmath}
\usepackage{amsthm}
\usepackage{xcolor} 
\usepackage{color}
\usepackage{framed}
\usepackage[hyphens]{url} 
\usepackage[Sonny]{fncychap}
\usepackage{fancyhdr}
\usepackage{cite}
\usepackage{slashed}
\usepackage{pstricks}
\usepackage{axodraw4j}
\usepackage{graphicx}
\usepackage{multirow}

\newcommand{\f}[2]{\frac{#1}{#2}}
\newcommand{\sss}[1]{\scriptscriptstyle#1}
\newcommand{\vv}[2]{\left( \begin{array}{c} #1 \\ #2  \end{array} \right)}
\newcommand{\bea}{\begin{eqnarray}}
\newcommand{\eea}{\end{eqnarray}}
\newcommand{\be}{\begin{equation}}
\newcommand{\ee}{\end{equation}}
\newcommand{\ba}{\begin{align}}
\newcommand{\ea}{\end{align}}
\newcommand{\beas}{\begin{eqnarray*}}
\newcommand{\eeas}{\end{eqnarray*}}
\newcommand{\bes}{\begin{equation*}}
\newcommand{\ees}{\end{equation*}}
\newcommand{\bas}{\begin{align*}}
\newcommand{\eas}{\end{align*}}

\newcommand{\gs}{g_{\scriptscriptstyle{s}}}
\newcommand{\yt}{y_{\scriptscriptstyle{t}}}

\newcommand{\als}{\alpha_{\scriptscriptstyle{s}}}

\newcommand{\lb}{\left(}
\newcommand{\rb}{\right)}

\definecolor{bluemar}{rgb}{0,0,.5}
\definecolor{redmar}{rgb}{.8,0,0}
\definecolor{greenmar}{rgb}{0,.5,0}

\title{Beta-function for the Higgs self-interaction in the Standard Model at three-loop level}

\ShortTitle{Beta-function for the Higgs self-interaction in the Standard Model at three-loop level}

\author{\speaker{Max ZOLLER}\\
        Karlsruhe Institute of Technology\\
        E-mail: \email{max.zoller@kit.edu}}

\abstract{
The discovery of a Higgs particle \cite{ATLAS:2012ae,Chatrchyan:2012tx} has triggered numerous theoretical and experimental investigations
concerning its production and decay rates and has led to interesting results concerning its interaction with fermions and gauge bosons.
The self-interaction $\lambda$ of the Standard Model Higgs boson is particularly important due to its close connection with the stability of the SM vacuum.
In this talk precision calculations for the evolution of this crucial coupling are presented
and their impact on the question of vacuum stability is analysed. We also compare the theoretical
precision resulting from the calculation of three-loop $\beta$-functions to the experimental uncertainties
stemming from key parameters, such as the top mass, the Higgs mass and the strong coupling, and to the theoretical uncertainties
introduced by the matching of experimental data to parameters in the theoretically favoured 
$\overline{\text{MS}}$ renormalization scheme.
}

\FullConference{The European Physical Society Conference on High Energy Physics \\
		 18-24 July, 2013\\
		 Stockholm, Sweden}

\begin{document}

\section{Introduction}
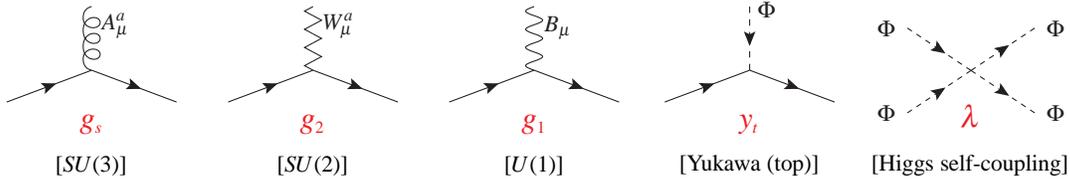
\begin{figure}[!h]\begin{center}
\scalebox{0.8}{
\begin{picture}(80,50) (0,0)
\ArrowLine(0,0)(40,15)
\ArrowLine(40,15)(80,0)
\Gluon(40,15)(40,45){4}{3.5}
 \Text(45,30)[lb]{{\Black{$A^a_\mu$}}}
 \Text(40,-15)[cb]{\Large{\Red{$g_{\sss{s}}$}}}
  \Text(40,-35)[cb]{[$SU$(3)]}
\end{picture} \qquad 
\begin{picture}(80,50) (0,0)
\ArrowLine(0,0)(40,15)
\ArrowLine(40,15)(80,0)
\ZigZag(40,15)(40,45){4}{3.5}
 \Text(45,30)[lb]{{\Black{$W^a_\mu$}}}
 \Text(40,-15)[cb]{\Large{\Red{$g_{\sss{2}}$}}}
  \Text(40,-35)[cb]{[$SU$(2)]}
\end{picture} \qquad 
\begin{picture}(80,50) (0,0)
\ArrowLine(0,0)(40,15)
\ArrowLine(40,15)(80,0)
\Photon(40,15)(40,45){4}{3.5}
 \Text(45,30)[lb]{{\Black{$B_\mu$}}}
 \Text(40,-15)[cb]{\Large{\Red{$g_{\sss{1}}$}}}
  \Text(40,-35)[cb]{[$U$(1)]}
\end{picture}\qquad
\begin{picture}(80,50) (0,0)
\ArrowLine(0,0)(40,15)
\ArrowLine(40,15)(80,0)
\DashArrowLine(40,45)(40,15){3}
 \Text(44,40)[lb]{{\Black{$\Phi$}}}
 \Text(40,-15)[cb]{\Large{\Red{$\yt$}}}
  \Text(40,-35)[cb]{[Yukawa (top)]}
\end{picture} \qquad
\begin{picture}(80,50) (0,-15)
\DashArrowLine(10,20)(40,0){3}
\DashArrowLine(10,-20)(40,0){3}
\DashArrowLine(40,0)(70,-20){3}
\DashArrowLine(40,0)(70,20){3}
 \Text(0,-20)[cc]{{\Black{$\Phi$}}}
 \Text(80,-20)[cc]{{\Black{$\Phi$}}}
 \Text(0,20)[cc]{{\Black{$\Phi$}}}
 \Text(80,20)[cc]{{\Black{$\Phi$}}}
 \Text(40,-27)[cb]{\Large{\Red{$\lambda$}}}
  \Text(40,-50)[cb]{[Higgs self-coupling]}
\end{picture}}
\vspace{0.5cm}\end{center}
\caption{SM interactions} \label{SMinter} 
\end{figure}

The Standard Model (SM) of particle physics is an $SU(3)\times SU(2) \times U(1)$
gauge theory describing the interactions of fermions through the exchange of gauge bosons.
In addition a scalar $SU(2)$ doublet is introduced which aquires a vacuum expectation value (VEV)
at the electroweak scale and produces the Higgs field and three Goldstone bosons. The fermion
masses and the Higgs-fermion interaction are described by the Yukawa sector of the SM and
the Higgs self-interaction is introduced to the Lagrangian in the Higgs potential which classically reads
\be V(\Phi)=\left(m^2\, \Phi^\dagger\Phi+\lambda\, \lb\Phi^\dagger\Phi\rb^2\right),\qquad
\Phi=\vv{\Phi_1}{\Phi_2}\underrightarrow{\text{SSB}} \vv{\Phi^+}{\f{1}{\sqrt{2}}(v+H-i\chi)}. \ee
The strength of each interaction is given by a coupling constant (see Fig. \ref{SMinter}).\footnote{Except
for $\yt$ the Yukawa interactions $y_b,\,y_c,\,\ldots$ can be neglected due to their smallness. The same
applies to the off-diagonal entries of the CKM-matrix.}

In the absence of physics beyond the SM at the LHC so far it is conceivable to
extrapolate the SM to high energies, eventually even up to the Planck scale. A problem which arises
as a consequence of radiative corrections is the possibility of a second minimum in the effective Higgs potential, 
lower than the one at the electroweak scale,
which would result in an unstable or metastable electroweak vacuum state.

The effective potential \cite{PhysRevD.7.1888} is affected by the self-interactions 
of the scalar fields as well as the interactions of the scalar fields with all other fields.
Hence it depends on all the SM couplings evolved from some initial scale, e.g. the top pole mass $M_t$,
up to the upper limit $\Lambda$ of the validity of the theory, e.g. $\Lambda=M_{Planck}$.

\begin{figure}[!h]
\begin{tabular}{ccc}
\includegraphics[height=3.3cm]{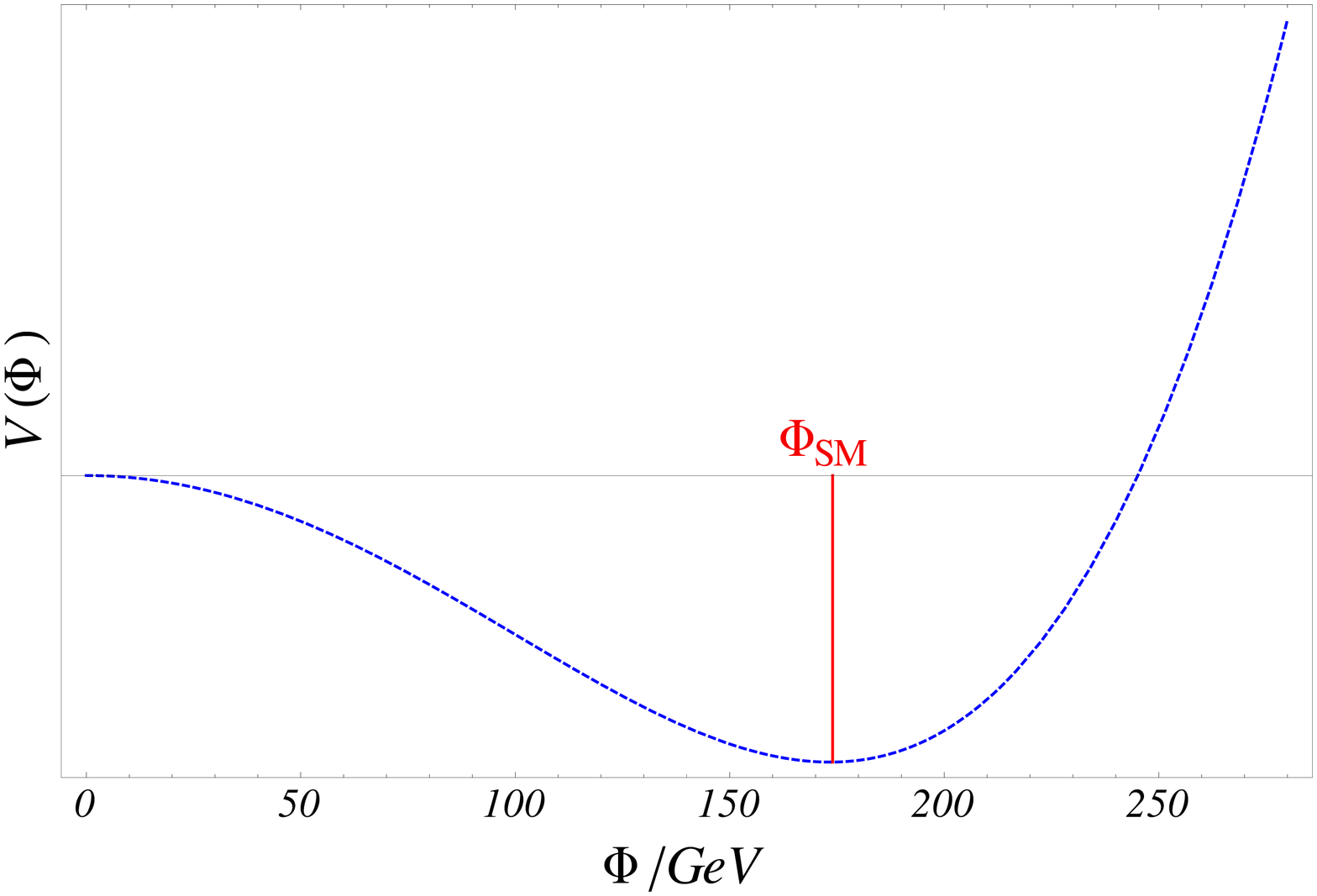} &
\includegraphics[height=3.4cm]{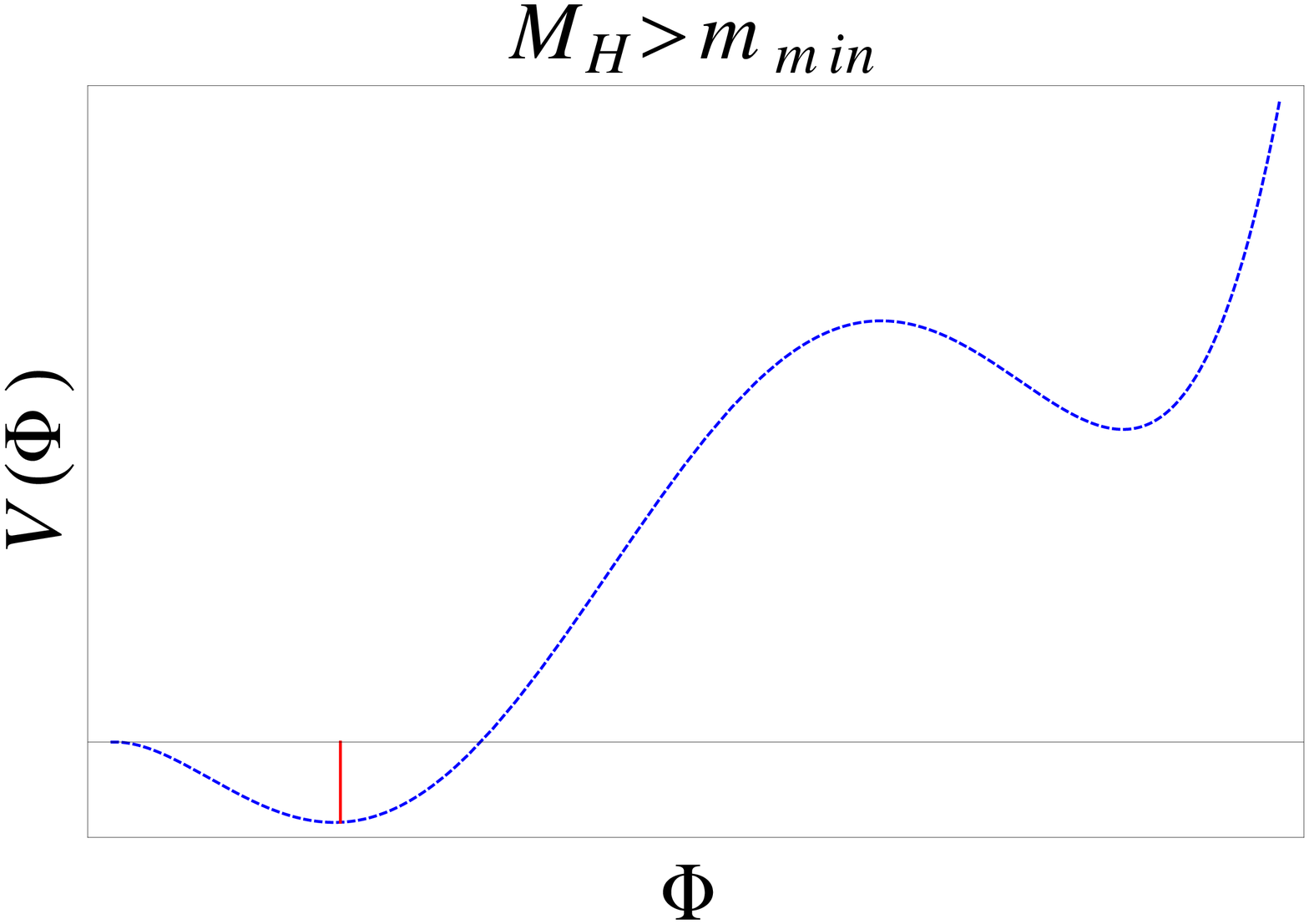} &
\includegraphics[height=3.4cm]{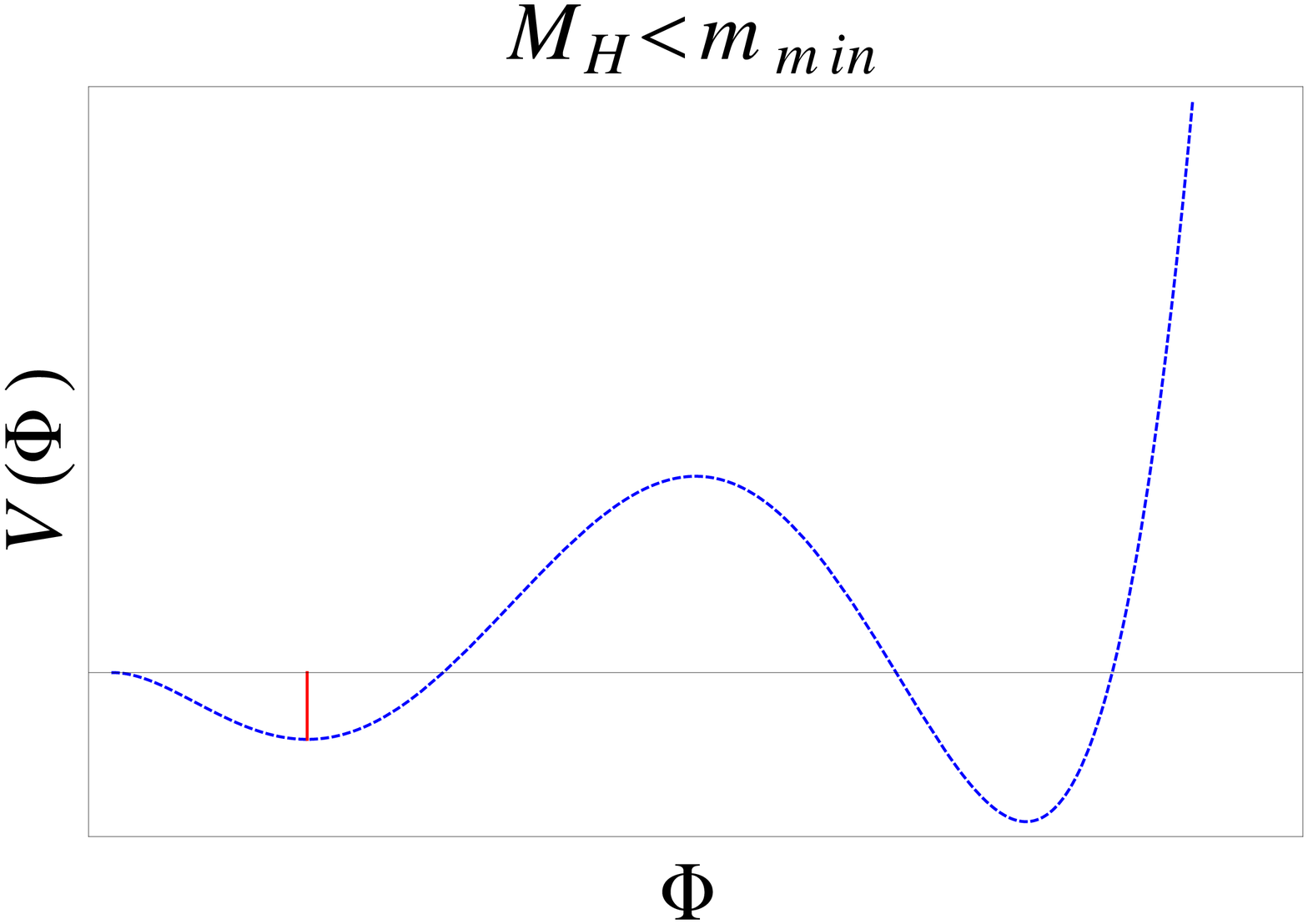}\\
classical situation & stable & unstable/metastable\\ (tree level)&
\fbox{$\lambda(\mu)>0\; \forall\; \mu\leq\Lambda$} &
\fbox{$\exists\; \mu\leq\Lambda:\; \lambda(\mu)<0$} 
\end{tabular}
\caption{Classical and effective Higgs potential} \label{Veffective} 
\end{figure}

Generic shapes of the classical Higgs potential
and of the effective potential are shown in Fig. \ref{Veffective} for the cases
of a Higgs mass larger and smaller than a critical value $m_{min}$, the minimal stability bound\footnote{There
is also an upper bound $m_{max}$ on the Higgs mass stemming from the requirement that no Landau pole appears 
at energies $\mu\leq \Lambda$.} (see also \cite{Bezrukov:2012sa}).
It has been demonstrated that the stability of the SM vacuum is in good approximation
equivalent to the question whether $\lambda$ stays positive up to the scale $\Lambda$
\cite{Altarelli1994141,Cabibbo:1979ay,Ford:1992mv}.
For a detailed discussion of the vacuum stability problem in the SM see
\cite{Buttazzo:2013uya,Masina:2012tz,Bezrukov:2012sa,Degrassi:2012ry,Zoller:2012cv,Chetyrkin:2012rz,EliasMiro:2011aa,Holthausen:2011aa}.

\section{The  three-loop $\beta$-function for the Higgs self-interaction}
The evolution of the Higgs self-coupling $\lambda$ is described by the $\beta$-function
\be \beta_{\lambda}(\lambda,\yt,\gs,g_2,g_1,\ldots)=\mu^2\f{d}{d\mu^2} \lambda(\mu), \label{betahiggs} \ee
which is a power series in all couplings of the SM and has recently been computed at three-loop level 
\cite{Chetyrkin:2012rz,Chetyrkin:2013wya,Bednyakov:2013eba}.
During the last two years the $\beta$-functions for the gauge \cite{PhysRevLett.108.151602,Mihaila:2012pz,Bednyakov:2012rb}
and Yukawa \cite{Chetyrkin:2012rz,Bednyakov:2012en} couplings have been calculated at three-loop level as well.
In order to determine the evolution of $\lambda$ we need to solve the coupled system of differential equations
\be \beta_{X}(\lambda,\yt,\gs,g_2,g_1)=\mu^2\f{d}{d\mu^2} X(\mu),\quad X \in \left\{\lambda,\yt,\gs,g_2,g_1\right\}. \label{betasystem} \ee
Furthermore, in order to solve eq.~\eqref{betasystem} an initial condition
for each coupling has to be given. One possible choice is to take their values at the scale of the top mass $M_t$.
As the $\beta$-functions have been calculated in the $\overline{\text{MS}}$-scheme but parameters like the top and Higgs mass
are (in good approximation) determined on-shell by experiments\footnote{This is particularly justified at a future linear
$e^+e^-$ collider, where the top mass for instance will be measured at the production threshold.}, we have to use matching relations 
between on-shell and $\overline{\text{MS}}$ parameters at two-loop level \cite{Buttazzo:2013uya,Jegerlehner:2012kn,Bezrukov:2012sa,
Espinosa:2007qp,Hempfling:1994ar,Sirlin1986389}.
For the pole masses $M_t=173.07$ GeV and $M_H=125.9$ GeV and $\als(M_Z)=0.1184$ \cite{PDG} we find:
\begin{table}[!h] \begin{center}
 \begin{tabular}{|l|l|}
  \hline
  $\gs(M_t)$ & $1.1667$\\
  $g_2(M_t)$ & $0.6483$\\
  $g_1(M_t)$ & $0.3587$\\
  $\yt(M_t)$ & $0.93543\pm 0.00050_{\text{(th,match)}}$\\
  $\lambda(M_t)$ & $0.12761\pm 0.00030_{\text{(th,match)}}$\\
    \hline
 \end{tabular} \end{center} \label{initialcond} \caption{Values for the SM couplings in the $\overline{\text{MS}}$-scheme at $\mu=M_t$ with 
the theoretical uncertainties for $\yt$ and $\lambda$ stemming from the matching procedure\cite{Buttazzo:2013uya}.}
\end{table} \newline

\section{The evolution of $\lambda$}
\begin{figure}[!h]
 \includegraphics[width=\linewidth]{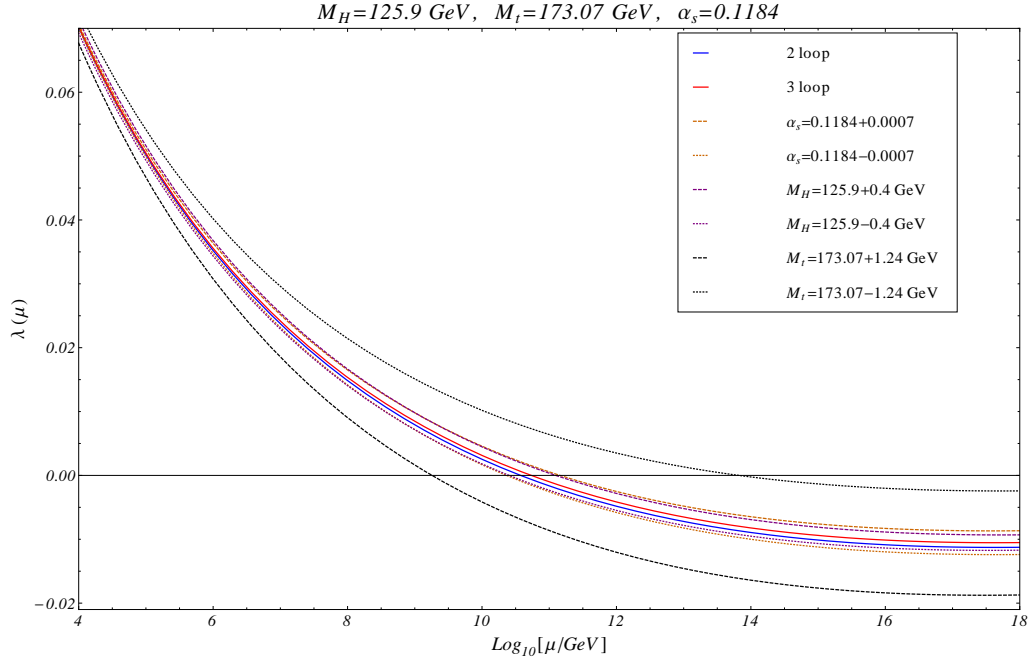}
\caption{Evolution of $\lambda$: experimental uncertainties} \label{lambda_evolution_Exp} 
\end{figure}
Using the three-loop $\beta$-functions for $\lambda,\yt,\gs,g_2$ and $g_1$ as well as the initial conditions from Tab.~\ref{initialcond}
we can plot the evolution of $\lambda$ up to the Planck scale \mbox{$M_{Planck}\sim10^{18}$ GeV}. For the 
experimental input parameters $M_t=173.07$ GeV, $M_H=125.9$ GeV and $\als(M_Z)=0.1184$ we find
the red curve in Fig.~\ref{lambda_evolution_Exp} which shows that $\lambda$ becomes negative at about $10^{11}$ GeV.
If we compare this to the evolution of $\lambda$ using only two-loop $\beta$-functions for the SM couplings (blue curve), we only find a
small deviation. The difference between these two curves can be interpreted as a measure for the theoretical uncertainty stemming from
the truncation of the perturbation series in the calculation of $\beta$-functions.
In contrast the experimental uncertainties are significantly larger. The dashed (dotted) lines show the behaviour of $\lambda$ evolved
using three-loop $\beta$-functions but with $M_t$, $M_H$ and $\als$ increased (decreased) by one standard deviation (values from \cite{PDG}).
The uncertainties originating from the experimental values for $\als$ and $M_H$ are roughly of the same size and a factor $2-3$ larger than
the difference between the two-loop and three-loop curves at the Planck scale. In contrast, the uncertainty stemming from the top mass
measurement is about an order of magnitude larger than the theoretical one at $\mu\sim 10^{18}$ GeV. 
It is worthy of note that if we shift all three experimental input parameters by one standard deviation in the more stable direction,
i.e. $M_t=(173.07-1.24)$ GeV, $M_H=(125.9+0.4)$ GeV and $\als(M_Z)=(0.1184+0.0007)$, 
we find $\lambda(\mu=10^{18} \text{GeV})\approx 0.00065\; > 0$ and hence a stable vacuum up to the Planck scale.

\begin{figure}[!h] \begin{center}
\includegraphics[width=0.8\linewidth]{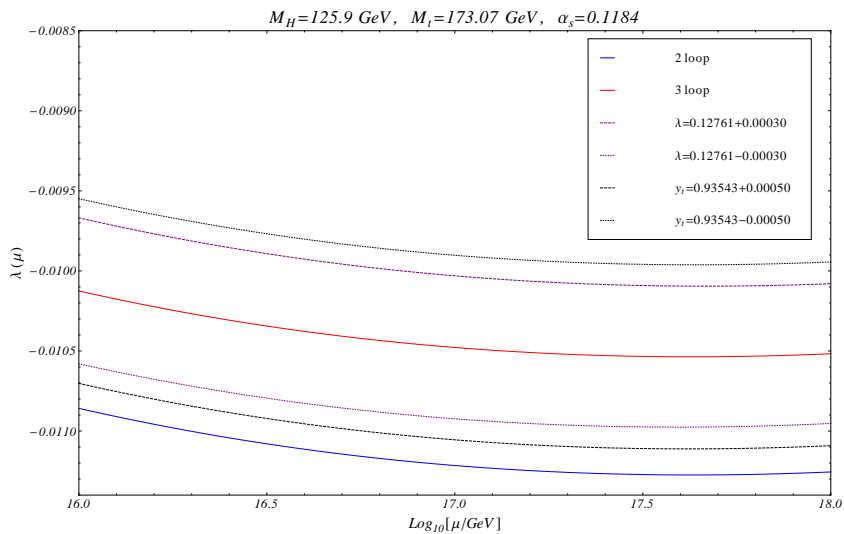} \end{center}
\caption{Evolution of $\lambda$: matching uncertainties} \label{lambda_evolution_Mat} 
\end{figure}

It is interesting to compare these uncertainties to the ones introduced by the matching procedure (see Table 1), i.e. 
by truncating the perturbation series in the matching formulas. In Fig.~\ref{lambda_evolution_Mat} the experimental input parameters are
fixed to $M_t=173.07$ GeV, $M_H=125.9$ GeV and $\als(M_Z)=0.1184$ and the $\lambda$-curves derived from two-loop (blue) and three-loop (red)
$\beta$-functions are given again. We focus on the region $\mu=10^{16}$ to $10^{18}$ GeV where the distances between different lines
are largest. The purple curves mark the uncertainty band due to the matching of the on-shell parameters to $\lambda$ in the 
$\overline{\text{MS}}$-scheme and the black curves describe the uncertainty originating from the matching to $\yt$ in the
$\overline{\text{MS}}$-scheme.
This plot clearly shows that all theoretical errors are roughly of the same size
and considerably smaller than the experimental ones.

\section{Conclusion}
The stability of the electroweak vacuum state is an interesting and fundamental problem in the SM framework.
Although it looks as if - in the absence of new physics - a metastable scenario is the most likely,
the present uncertainties do not allow for a definitive answer. The analysis presented in this talk shows that
the theoretical uncertainties are well under control due to the calculation of three-loop $\beta$-functions for the Higgs
self-interaction and all other SM couplings as well as due to improved precision in the matching relations between
on-shell and $\overline{\text{MS}}$-parameters.
A more precise measurement of the experimental input parameters, especially the top mass, will hopefully
lead to a clarification of this issue in the future.

\section*{Acknowledgments}
I thank my collaborator K. G. Chetyrkin for invaluable discussions
and J. K\"uhn for his support and useful comments. 
This work has been supported by the Deutsche Forschungsgemeinschaft in the
Sonderforschungsbereich/Transregio SFB/TR-9 ``Computational Particle
Physics'' and the Graduiertenkolleg ``Elementarteilchenphysik
bei h\"ochsten Energien und h\"ochster Pr\"azission''


\begin{thebibliography}{99}
\scriptsize

\bibitem{ATLAS:2012ae}
{\bf ATLAS} Collaboration, G.~Aad {\em et al.}, Phys. Lett. {\bf B710}
  (2012)  49--66,
\href{http://arxiv.org/abs/1202.1408}{{\tt arXiv:1202.1408 }}.

\bibitem{Chatrchyan:2012tx}
{\bf CMS} Collaboration, S.~Chatrchyan {\em et al.}, Phys.\ Lett.\ B {\bf 710}, 26 (2012),
\href{http://arxiv.org/abs/1202.1488}{{\tt arXiv:1202.1488 }}.

\bibitem{PhysRevD.7.1888}
S.~Coleman and E.~Weinberg, 
  \href{http://dx.doi.org/10.1103/PhysRevD.7.1888}{Phys. Rev. D {\bf 7} (1973)
  1888--1910}.

\bibitem{Bezrukov:2012sa}
F.~Bezrukov, M.~Y. Kalmykov, B.~A. Kniehl, and M.~Shaposhnikov, JHEP {\bf 1210}, 140 (2012),
\href{http://arxiv.org/abs/1205.2893}{{\tt
  arXiv:1205.2893 }}.

\bibitem{Altarelli1994141}
G.~Altarelli and G.~Isidori, 
  \href{http://dx.doi.org/10.1016/0370-2693(94)91458-3}{Physics Letters B {\bf
  337} (1994) no. 1-2, 141--144}.

\bibitem{Cabibbo:1979ay}
N.~Cabibbo, L.~Maiani, G.~Parisi, and R.~Petronzio, 
\href{http://dx.doi.org/10.1016/0550-3213(79)90167-6}{Nucl. Phys. {\bf B158}
  (1979)  295--305}.

\bibitem{Ford:1992mv}
C.~Ford, D.~Jones, P.~Stephenson, and M.~Einhorn, 
  \href{http://dx.doi.org/10.1016/0550-3213(93)90206-5}{Nucl. Phys. {\bf B395}
  (1993)  17--34},
\href{http://arxiv.org/abs/hep-lat/9210033}{{\tt arXiv:hep-lat/9210033
  }}.

\bibitem{Buttazzo:2013uya}
D.~Buttazzo, G.~Degrassi, P.~P. Giardino, G.~F. Giudice, F.~Sala, {\em et al.},
\href{http://arxiv.org/abs/1307.3536}{{\tt arXiv:1307.3536 }}.

\bibitem{Masina:2012tz}
I.~Masina, Phys.\ Rev.\ D {\bf 87}, no. 5, 053001 (2013),
\href{http://arxiv.org/abs/1209.0393}{{\tt arXiv:1209.0393 }}.

\bibitem{Degrassi:2012ry}
G.~Degrassi, S.~Di~Vita, J.~Elias-Miro, J.~R. Espinosa, G.~F. Giudice, {\em et
  al.}, 
  \href{http://dx.doi.org/10.1007/JHEP08(2012)098}{JHEP {\bf 1208} (2012)
  098},
\href{http://arxiv.org/abs/1205.6497}{{\tt arXiv:1205.6497 }}.

\bibitem{Zoller:2012cv}
M.~Zoller, 
\href{http://arxiv.org/abs/1209.5609}{{\tt arXiv:1209.5609 }}.

\bibitem{Chetyrkin:2012rz}
K.~Chetyrkin and M.~Zoller, 
  \href{http://dx.doi.org/10.1007/JHEP06(2012)033}{JHEP {\bf 1206} (2012)
  033}, \href{http://arxiv.org/abs/1205.2892}{{\tt arXiv:1205.2892 }}.

\bibitem{EliasMiro:2011aa}
J.~Elias-Miro, J.~R. Espinosa, G.~F. Giudice, G.~Isidori, A.~Riotto, {\em et
  al.},  \href{http://dx.doi.org/10.1016/j.physletb.2012.02.013}{Phys. Lett.
  {\bf B709} (2012)  222--228},
\href{http://arxiv.org/abs/1112.3022}{{\tt arXiv:1112.3022 }}.

\bibitem{Holthausen:2011aa}
M.~Holthausen, K.~S. Lim, and M.~Lindner, 
  \href{http://dx.doi.org/10.1007/JHEP02(2012)037}{JHEP {\bf 1202} (2012)
  037},
\href{http://arxiv.org/abs/1112.2415}{{\tt arXiv:1112.2415 }}.

\bibitem{Chetyrkin:2013wya}
K.~Chetyrkin and M.~Zoller, 
  \href{http://dx.doi.org/10.1007/JHEP04(2013)091}{JHEP {\bf 1304} (2013)
  091},
\href{http://arxiv.org/abs/1303.2890}{{\tt arXiv:1303.2890 }}.

\bibitem{Bednyakov:2013eba}
A.~Bednyakov, A.~Pikelner, and V.~Velizhanin, Nucl.\ Phys.\ B {\bf 875}, 552 (2013),
\href{http://arxiv.org/abs/1303.4364}{{\tt arXiv:1303.4364 }}.

\bibitem{PhysRevLett.108.151602}
L.~N. Mihaila, J.~Salomon, and M.~Steinhauser, 
  \href{http://dx.doi.org/10.1103/PhysRevLett.108.151602}{Phys. Rev. Lett. {\bf
  108} (2012)  151602}.

\bibitem{Mihaila:2012pz}
L.~N. Mihaila, J.~Salomon, and M.~Steinhauser, Phys.\ Rev.\ D {\bf 86}, 096008 (2012), \href{http://arxiv.org/abs/1208.3357}{{\tt
  arXiv:1208.3357 }}.

\bibitem{Bednyakov:2012rb}
A.~Bednyakov, A.~Pikelner, and V.~Velizhanin,  JHEP {\bf 1301} (2013)  017,
\href{http://arxiv.org/abs/1210.6873}{{\tt arXiv:1210.6873 }}.

\bibitem{Bednyakov:2012en}
A.~Bednyakov, A.~Pikelner, and V.~Velizhanin, Phys.\ Lett.\ B {\bf 722}, 336 (2013),
\href{http://arxiv.org/abs/1212.6829}{{\tt arXiv:1212.6829 }}.

\bibitem{PhysRevLett.30.1343}
D.~J. Gross and F.~Wilczek, \href{http://dx.doi.org/10.1103/PhysRevLett.30.1343}{Phys. Rev.
  Lett. {\bf 30} (1973)  1343--1346}.

\bibitem{PhysRevLett.30.1346}
H.~D. Politzer, 
  \href{http://dx.doi.org/10.1103/PhysRevLett.30.1346}{Phys. Rev. Lett. {\bf
  30} (1973)  1346--1349}.

\bibitem{Jones1974531}
D.~Jones,
  \href{http://dx.doi.org/10.1016/0550-3213(74)90093-5}{Nuclear Physics B {\bf
  75} (1974) no.~3, 531--538}.

\bibitem{Tarasov:1976ef}
O.~Tarasov and A.~Vladimirov,
Sov.J.Nucl.Phys. {\bf 25} (1977)  585.

\bibitem{PhysRevLett.33.244}
W.~E. Caswell, 
  \href{http://dx.doi.org/10.1103/PhysRevLett.33.244}{Phys.Rev.Lett. {\bf 33}
  (1974)  244--246}.

\bibitem{Egorian:1978zx}
E.~Egorian and O.~Tarasov, 
Teor.Mat.Fiz. {\bf 41} (1979)  26--32.

\bibitem{PhysRevD.25.581}
D.~R.~T. Jones, \href{http://dx.doi.org/10.1103/PhysRevD.25.581}{Phys. Rev. D
  {\bf 25} (1982)  581--582}.

\bibitem{Fischler:1981is}
M.~S. Fischler and C.~T. Hill, 
\href{http://dx.doi.org/10.1016/0550-3213(81)90517-4}{Nucl.Phys. {\bf B193}
  (1981)  53}.

\bibitem{Fischler1982385}
M.~Fischler and J.~Oliensis, 
  \href{http://dx.doi.org/10.1016/0370-2693(82)90695-5}{Phys. Lett. B {\bf 119}
  (1982) no.~4, 385--386}.

\bibitem{Jack1985472}
I.~Jack and H.~Osborn, \href{http://dx.doi.org/10.1016/0550-3213(85)90088-4}{Nucl. Phys. B
  {\bf 249} (1985) no.~3, 472--506}.

\bibitem{Machacek198383}
M.~E. Machacek and M.~T. Vaughn, 
  \href{http://dx.doi.org/10.1016/0550-3213(83)90610-7}{Nucl. Phys. B {\bf 222}
  (1983) no. 1, 83--103}.

\bibitem{Machacek1984221}
M.~E. Machacek and M.~T. Vaughn, 
  \href{http://dx.doi.org/10.1016/0550-3213(84)90533-9}{Nucl. Phys. B {\bf 236}
  (1984) no. 1, 221--232}.

\bibitem{Machacek198570}
M.~E. Machacek and M.~T. Vaughn, 
  \href{http://dx.doi.org/10.1016/0550-3213(85)90040-9}{Nucl. Phys. B {\bf 249}
  (1985) no. 1, 70--92}.

\bibitem{2loopbetayukawa}
M.-x. Luo and Y.~Xiao, 
  \href{http://dx.doi.org/10.1103/PhysRevLett.90.011601}{Phys. Rev. Lett. {\bf
  90} (2003)  011601},
\href{http://arxiv.org/abs/hep-ph/0207271}{{\tt arXiv:hep-ph/0207271
  }}.

\bibitem{Ford:1992pn}
C.~Ford, I.~Jack, and D.~Jones,\href{http://dx.doi.org/10.1016/0550-3213(92)90165-8}{Nucl.Phys.
  {\bf B387} (1992)  373--390},
\href{http://arxiv.org/abs/hep-ph/0111190}{{\tt arXiv:hep-ph/0111190
  }}.

\bibitem{Curtright:1979mg}
T.~Curtright, 
\href{http://dx.doi.org/10.1103/PhysRevD.21.1543}{Phys.Rev. {\bf D21} (1980)
  1543}.

\bibitem{Jones:1980fx}
D.~Jones, 
\href{http://dx.doi.org/10.1103/PhysRevD.22.3140}{Phys.Rev. {\bf D22} (1980)
  3140--3141}.

\bibitem{Tarasov:1980au}
O.~Tarasov, A.~Vladimirov, and A.~Y. Zharkov, 
\href{http://dx.doi.org/10.1016/0370-2693(80)90358-5}{Phys.Lett. {\bf B93}
  (1980)  429--432}.

\bibitem{3loopbetaqcd}
S.~Larin and J.~Vermaseren, 
  \href{http://dx.doi.org/10.1016/0370-2693(93)91441-O}{Phys. Lett. {\bf B303}
  (1993)  334--336},
\href{http://arxiv.org/abs/hep-ph/9302208}{{\tt arXiv:hep-ph/9302208
  }}.

\bibitem{Steinhauser:1998cm}
M.~Steinhauser,  \href{http://dx.doi.org/10.1103/PhysRevD.59.054005}{Phys.Rev.
  {\bf D59} (1999)  054005},
\href{http://arxiv.org/abs/hep-ph/9809507}{{\tt arXiv:hep-ph/9809507
  }}.

\bibitem{Pickering:2001aq}
A.~Pickering, J.~Gracey, and D.~Jones, 
  \href{http://dx.doi.org/10.1016/S0370-2693(01)00624-4}{Phys.Lett. {\bf B510}
  (2001)  347--354},
\href{http://arxiv.org/abs/hep-ph/0104247}{{\tt arXiv:hep-ph/0104247
  }}.

\bibitem{Jegerlehner:2012kn}
F.~Jegerlehner, M.~Y. Kalmykov, and B.~A. Kniehl, \href{http://dx.doi.org/10.1016/j.physletb.2013.04.012}{Phys.Lett.
  {\bf B722} (2013)  123--129},
\href{http://arxiv.org/abs/1212.4319}{{\tt arXiv:1212.4319 }}.

\bibitem{Espinosa:2007qp}
J.~Espinosa, G.~Giudice, and A.~Riotto, 
  \href{http://dx.doi.org/10.1088/1475-7516/2008/05/002}{JCAP {\bf 0805} (2008)
   002},
\href{http://arxiv.org/abs/0710.2484}{{\tt arXiv:0710.2484 }}.

\bibitem{Hempfling:1994ar}
R.~Hempfling and B.~A. Kniehl, 
  \href{http://dx.doi.org/10.1103/PhysRevD.51.1386}{Phys.Rev. {\bf D51} (1995)
  1386--1394},
\href{http://arxiv.org/abs/hep-ph/9408313}{{\tt arXiv:hep-ph/9408313
  }}.

\bibitem{Sirlin1986389}
A.~Sirlin and R.~Zucchini, 
  \href{http://dx.doi.org/10.1016/0550-3213(86)90096-9}{Nucl. Phys. B {\bf 266}
  (1986) no.~2, 389--409}.

\bibitem{PDG}
K.~Nakamura {\em et al.},  J. Phys. {\bf
  G} (2010) no.~37, 075021.

\end{thebibliography}

\providecommand{\href}[2]{#2}\begingroup\raggedright
\endgroup

\end{document}